\documentclass[pra,aps,twocolumn,showpacs,superscriptaddress]{revtex4-1}
\usepackage{graphicx}
\usepackage{multirow}
\usepackage{amsmath}
\usepackage{amssymb}
\usepackage{amsthm}
\usepackage{eucal}
\usepackage{color}
\usepackage{bm}
\usepackage{upgreek}
\usepackage{mathrsfs}
\usepackage{latexsym}
\usepackage{float}
\usepackage{hyperref}
\usepackage{setspace}
\usepackage{amsmath}
\usepackage{amssymb}
\usepackage{bm}

\usepackage[english]{babel}
\usepackage{blindtext}
\newcommand{\s}{\bf{s}}
\newcommand{\h}{\hbar}
\newtheorem{proposition}{Proposition}
\newtheorem{theorem}{Theorem}
\newcommand{\hr}[1]{\hat{\rho}_{#1}}
\newcommand{\sca}{\Gamma_{{\rm s},a}}
\begin{document}
\title{Scaling maps of s-ordered quasiprobabilities are either nonpositive or completely positive}
\author{J. Solomon Ivan}
\email{solomonivan@iist.ac.in}
\affiliation{Department of Physics, 
Indian Institute of Space Science and Technology, Trivandrum 
695 547.}
\author{Krishna Kumar Sabapathy}
\email{krishnakumar.sabapathy@gmail.com}
\affiliation{Centre for Quantum Information and Communication, Ecole polytechnique de Bruxelles, CP 165, Universit\'{e} Libre de Bruxelles, 1050 Brussels, Belgium.}
\affiliation{F\'{i}sica Te\`{o}rica: Informaci\'{o} i Fen\`{o}mens Qu\`{a}ntics, Universitat Aut\`{o}noma de Barcelona, ES-08193 Bellaterra (Barcelona), Spain.}
\author{R. Simon}
\email{simon@imsc.res.in}
\affiliation{Optics \& Quantum Information Group, The Institute of Mathematical
  Sciences, HBNI, C.I.T Campus, Tharamani, Chennai 600 113, India.}
\begin{abstract} 
Continuous-variable systems in quantum theory can be fully described 
through any one of the ${\rm s}$-ordered family of quasiprobabilities
$\Lambda_{\rm s}(\alpha)$, ${\rm s} \in [-1,1]$. We ask for what
values of $({\rm s}, a)$ is the scaling map $\Lambda_{\rm s}(\alpha) \rightarrow 
a^{-2} \Lambda_{\rm s}(a^{-1}\alpha)$ a positive map? Our analysis based on a 
duality we establish settles this issue (i) the scaling map
generically fails to be positive, showing that there is no useful
entanglement witness of the scaling type beyond the transpose map, and
(ii) in the two particular cases $({\rm s}=1, |a| \leq 1)$ and 
$({\rm s}=-1, |a| \geq 1)$, and only in these two non-trivial cases,  the map is not only  
positive but also completely positive as seen through the noiseless attenuator and amplifier channels. We also present a `phase diagram' for the behaviour
of the scaling maps in the ${\rm s}-a$ parameter space with regard to its positivity, obtained from the viewpoint of symmetric-ordered characteristic functions. This also sheds light on similar diagrams for the practically relevant attenuation and amplification maps with respect to the noise parameter, especially in the range below the complete-positivity (or quantum-limited) threshold.
\end{abstract}
\maketitle

\section{Introduction}
Completely  positive maps mathematically describe physical processes or quantum channels\,\cite{Sudarshan,Krausb,Choi,reff1,qdisccp}. 
Positive maps  which fail to
be completely positive cannot represent physical processes, but they 
play a  key role in the study of inseparability of mixed states as 
entanglement witnesses\,\cite{peres96, horodecki96,horodeckirmp}. 
The desirability of studying maps represented by uniform 
scaling of phase   space variables of continuous-variable systems, namely  $\hat{q}   \to  a  \hat{q}$,   $\hat{p}
\to a  \hat{p}$, $a \in {\mathbb R}\backslash\{0\}$ was suggested in Ref.\,\cite{Manko} in the hope  that {\em such a map might be positive but not completely positive}, and hence partial scaling (i.e., uniform phase space scaling on one party of a bipartite state) could
prove useful as an entanglement witness, `{\em generalizing}'\,\cite{Manko} the partial momentum 
reversal or partial transpose criterion\,\cite{Simon} for separability. 
These authors pointed to the interesting construct that this scaling of 
phase space could, alternatively, be viewed 
as scaling of the  Planck constant: $\hbar \to a^2\hbar$.
(We shall comment on this towards the end of the paper). 
Note that the signature of the scale parameter $a$ can be changed simply 
through a natural unitary evolution of the mode through half a 
period.
 
Though this  uniform scaling is a linear  transformation
 at the operator level,  it  is  not
 canonical for $|a| \ne 1$, and  hence  cannot  be implemented as  a linear,
unitary transformation  on Hilbert space  vectors.
We therefore introduce abstract scaling maps $\sca$ whose action on the density operators can be represented at the level of ${\rm s}$-ordered quasiprobabilities $\Lambda_{\rm \s}(q,\,p)$ in the following manner\,:
\begin{align}
\label{sca}
\sca : \Lambda_{\rm s}(q,p\,;\sca(\hr{})) = a^{-2} \Lambda_{\rm s}(a^{-1}q, a^{-1}p\,;\hr{}),
\end{align}
with ${\rm s} \in [-1,1]$, $a \in {\mathbb R}\backslash \{0\}$.

The Wigner distribution 
(${\rm \s} = 0$) was the choice of Ref.\,\cite{Manko} to implement the scaling transformation through 
\begin{align}
 W(q,\,p;\hr{})  \to 
W(q,\,p;\hr{}^{\, '}) =a^{-2}\, W(a^{-1}q,\,a^{-1} p;\hr{}).
\end{align}
  That this map preserves hermiticity
and normalization of density operators is transparent. It turns out that it does not
preserve the  nonnegativity property of density operators  (see Proposition \ref{prop2} and also Refs.\,\cite{narcowich,holevo16}), showing
that   the   scaling   map      defined in this manner   through   the   Wigner
quasiprobability is  not positive. The original expectation  of the authors of
Ref.\,\cite{Manko} thus turns out  to be  unfounded.

If a positive map  obtains for any other value of order parameter ${\rm \s}$,
then there are  two possibilities\,: (a) the
positive map may not  be completely 
positive,  in  which  case  it  will  be  useful  as  an  entanglement
witness; (b) if it turns out to be completely positive
it will correspond to a quantum channel. Indeed, since any scaling map ${\sca}$ transforms  Gaussian quasiprobability distributions  into  Gaussian
distributions, any completely positive scaling map  will  correspond to a  
bosonic Gaussian channel, a topic of considerable current  
interest\,\cite{holevo16,hw01,kraus10,robust,raulrmp,raulprl,nb,holevoext,raulnat,gio2,nbm,ncd,gio1,ntpg,caruso06,holevoebt}. 

The main purpose of the present work is to 
address the issue of positivity of scaling maps, a generalization of the one raised in Ref.\,\cite{Manko}, in a definitive manner in Sec. II.
In Sec. III we introduce a `phase diagram' for the behaviour of the scaling maps $\sca$ over the ${\rm s}-a$ parameter space with regard to its positivity. 
As an almost unintended  fall out of our analysis, we present a related phase diagram  for attenuation and amplification maps with respect to the noise parameter of the map, especially depicting the behaviour of these maps below the complete-positivity or quantum-limited noise threshold. We conclude in Sec. IV.

\section{Quasiprobabilities and scaling transformations} 
Quasiprobabilities  are  defined more  conveniently  in  terms of  the
nonhermitian   operators  $\hat{a},\,  \hat{a}^{\dagger}$   obeying  the
commutation relation $[ \hat{a},\, \hat{a}^{\dagger}] = 1$, than in terms
of their hermitian parts $\hat{q}, \hat{p}$ obeying $[\hat{q},\,
\hat{p}] = i$. In correspondence  with the relation $\hat{a} = (\hat{q}
+ i \hat{p}) / \sqrt{2}$ we will associate to every point $(q,\,p)$ in 
the phase space plane the complex number  $\alpha  =  (q +  i  p)  / \sqrt{2}$.  The
characteristic function $\chi_{{\rm \s}}(\xi;\hr{})$ of the {\bf s}-ordered 
quasiprobability
$\Lambda_{{\rm \s}}(\alpha;\hr{})$ associated with a  density operator  $\hat{\rho}$ is
defined through \cite{Cahill-Glauber}
\begin{align}
& \chi_{\rm \s} (\xi;\hr{}) = \exp{[\,\frac{\,1\,}{2}\, {\rm \s}\, |\xi|^2\,]}\, 
\chi_0 
(\xi;\hr{}),~~~ -1 \leq {\rm \s}
  \leq 1; \nonumber \\
& \chi_0(\xi;\hr{}) = \text{Tr}\, (\,\hat{\rho} {\cal D}(\xi)\,),~~~ {\cal D}(\xi) = 
\text{exp}\,{[\,\alpha
  \hat{a}^{\dagger} - \alpha^* \hat{a} \,]}. \label{0}
\end{align}
For every $-1 \leq {\rm \s} \leq 1$, the quasiprobability 
$\Lambda_{{\rm \s}} (\alpha;\hr{})$ and the associated characteristic function 
$\chi_{{\rm \s}} (\xi;\hr{})$ are related through the Fourier pair  
\begin{align}\label{fourier}
& \Lambda_{{\rm \s}}(\alpha;\hr{}) = \pi^{-1} \int \text{exp}\,{[\,\alpha \xi^* - 
\alpha^* \xi\,]}\,
\chi_{{\rm \s}}(\xi;\hr{})\,d^{\,2} \xi, \nonumber \\
&  \chi_{{\rm \s}}(\xi;\hr{}) = \pi^{-1} \int \text{exp}\,{[\,\xi \alpha^* - 
\xi^* 
\alpha\,]}\,
\Lambda_{\rm \s}(\alpha;\hr{})\,d^{\, 2} \alpha.
\end{align}
The familiar diagonal `weight' function $\phi$ \,\cite{Sudarshan63} or $P$-distribution 
\cite{Glauber63}, Wigner distribution $W$, 
and Husimi or $Q$-distribution correspond respectively to ${\rm 
s} = 1,\,0,\,-1$ \,\cite{Cahill-Glauber}. It is clear 
that the association between operators $\hat{\rho}$ 
and  quasiprobabilities  $\Lambda_{\rm \s}(\alpha;\hr{})$ (or characteristic 
functions $\chi_{{\rm \s}}(\xi;\hr{})$) is one-to-one invertible
for any $-1\le {\rm \s}\le 1$. While 
hermiticity of $\hat{\rho}$ translates into reality of $\Lambda_{\rm \s}(\alpha;\hr{})$ 
and the normalization ${\rm Tr}\,\hat{\rho} =1$ into the normalization
$\int \Lambda_{\rm \s}(\alpha;\hr{})\,d^{\,2}\alpha = 1$ [or equivalently, $\chi_{\rm \s}(\xi;\hr{})|_{0}=1$], 
positivity of $\hat{\rho}$ gets encoded in $\Lambda_{\rm \s}(\alpha;\hr{})$ in a more 
subtle manner\,\cite{Cahill-Glauber}. 

In light of Eqs.\,\eqref{sca} and \eqref{fourier}, we can rewrite the action of the scaling maps $\sca$ as
\begin{align}
&\Lambda_{\rm s}(\alpha;\sca(\hr{})) = a^{-2}\Lambda_{\rm s}(a^{-1}\alpha;\hr{}).
\label{eq5}
\end{align}
It is clear from the Fourier transform pair \eqref{fourier}  that 
the scaling transformation ${\sca}$ described through its action on {\bf s}-ordered quasiprobability (Eq. \eqref{eq5}) reads, 
when transcribed to the associated characteristic
function, as the transformation 
\begin{align}
\label{scacar}
\chi_{\rm s}(\xi;\sca(\hr{})) = \chi_{\rm s}(a\xi;\hr{}).
\end{align}
Also, it turns out to be useful for succeeding sections to introduce a `dual' scaling map $\widetilde{\Gamma}_{{\rm s},a}$ in the following way
\begin{align}
\label{dusca}
\widetilde{\Gamma}_{{\rm s},a} : \Lambda_{-\rm s}(\alpha;\widetilde{\Gamma}_{{\rm s},a}(\hr{})) = a^{2} \Lambda_{-{\rm s}}(a\alpha;\hr{}).
\end{align}
Note that while the action of the map ${\sca}$ is described at the level of $\Lambda_{\rm s}$ its dual map $\widetilde{\Gamma}_{{\rm s},a}$ is described at the level of $\Lambda_{-\rm s}$ with the scaling $a$ replaced by its inverse.

We begin our analysis of scaling maps by establishing an important duality between the pair of scaling maps $\sca$ and $\widetilde{\Gamma}_{{\rm s},a}$ with regard to positivity.  
\begin{proposition}[Duality]
 The scaling map $\Gamma_{{\rm s},a^{-1}}$ is positive if and only if the
dual map $\widetilde{\Gamma}_{{\rm s},a^{-1}}$  is positive.   
  \label{prop1}
  \end{proposition}
\noindent
{\em Proof}: Positivity of entities are often defined or described through 
the `company they
keep'. Thus an operator $\hat{\rho}$ is positive if and only if
$\text{Tr}\,(\hat{\rho} \hat{\rho}^{\,\prime} ) \geq 0$ for all positive 
operators $\hat{\rho}^{\, \prime}$. Transcription of this statement to the language of 
{\bf s}-ordered quasiprobabilities involves a dual pair of quasiprobabilities
$\Lambda_{\rm \s}(\alpha;\hr{})$ and $\Lambda_{\rm -\s}(\alpha;\hr{})$ and reads\,: 
$\Lambda_{{\rm \s}} (\alpha;\hr{})$ is a {\bf s}-ordered quasiprobability (i.e. it
corresponds to a positive operator) if and only if 
\begin{equation}\label{1}
\int \Lambda_{{\rm \s}} (\alpha;\hr{}) \Lambda_{-{\rm \s}} (\alpha;\hr{}^{\, '}) \,d^{\, 2} \alpha \geq 0,
\end{equation}
for every (--\,{\bf s})-ordered quasiprobability $ \Lambda_{-{\rm \s}} (\alpha;\hr{}^{\, '})$. 
It is in this sense that the {\bf s}-ordered quasiprobabilities 
$\Lambda_{\rm \s}(\alpha;\hr{})$ and the (--\,{\bf s})-ordered
quasiprobabilities $\Lambda_{\rm -\s}(\alpha;\hr{})$ are mutually dual. 
In particular, the quasiprobability of Wigner is self-dual, and this is the only
self-dual {\bf s}-ordered quasiprobability, while the quasiprobabilities $Q$ and $P$
are mutually dual.

Now, by definition, the map  $\Gamma_{{\rm s},a^{-1}}$ is  positive if and only if $ a^2 \Lambda_{{\rm \s}}(a\alpha;\hr{})$ is an {\bf s}-ordered  quasiprobability for every {\bf s}-ordered
quasiprobability $\Lambda_{{\rm \s}}(\alpha;\hr{})$. In view of \eqref{1}, the necessary and
sufficient condition for $\Gamma_{{\rm s},a^{-1}}$ to be positive is that 
\begin{equation}
\label{2} 
\int \Lambda_{{\rm \s}} (a\alpha;\hr{}) \Lambda_{-{\rm \s}} (\alpha;\hr{}^{\, '}) \,d^{\, 2} \alpha \geq 0,
\end{equation}
for every {\bf s}-ordered quasiprobability $\Lambda_{{\rm \s}} (\alpha;\hr{})$ and  
(--\,{\bf s})-ordered quasiprobability $\Lambda_{-{\rm \s}} (\alpha;\hr{}^{\, '})$. Since $a \neq 0$, the last 
stipulation 
\eqref{2} is the same as the condition that  
\begin{equation}
\int \Lambda_{{\rm \s}} (\alpha;\hr{}) \Lambda_{-{\rm \s}} (a^{-1} \alpha;\hr{}^{\, '}) \,d^{\, 2} 
\alpha \geq 0, 
\end{equation}
for every {\bf s}-ordered quasiprobability $\Lambda_{{\rm \s}} (\alpha;\hr{})$ and 
[(--\,{\bf s})-ordered
quasiprobability $\Lambda_{-{\rm \s}} (\alpha;\hr{}^{\, '})$]. But this condition precisely
constitutes the assertion that the scaling map $\widetilde{\Gamma}_{{\rm s},a^{-1}}$ 
is a positive map, and thus completes proof of
the proposition.\hfill$\blacksquare$

It is well known that the Gaussian function 
$\text{exp}\,{[\,-\,\frac{1}{2} b 
|\xi|^2\,]}$
qualifies to be the characteristic function of some {\bf s}-ordered
quasiprobability if and only if $ b \geq 1-{\rm \s} $. This is basically a
statement of the Heisenberg uncertainty principle. Indeed, saturation
of this inequality corresponds to the ground state of the oscillator
$\hat{\rho} = |0 \rangle \langle 0|$, and this is true for every 
$-1 \leq {\rm \s} \leq 1$. The case $ b > 1-{\rm \s}$ 
corresponds to thermal states with the temperature of the state being a 
monotone increasing function of
the difference $b - (1-{\rm \s}) = b +{\rm \s} -1$. 

Subjecting the ground state  
to the transformation $\Gamma_{{\rm s},a}$ one readily concludes, {\em for all} ${\rm \s} \ne 1$,  that a 
necessary condition for
this map to be positive is that $a \geq 1$. Note that {\em this requirement is
independent of the value of ${\rm \s}$} (excluding ${\rm \s} = 1$). Applying 
this requirement on the vacuum state to both the cases  
${\rm \s}$ and $-{\rm \s}$ 
we conclude, in view of the duality established
in Proposition \ref{prop1} that
\begin{proposition}
\label{prop2}
The scaling map $\sca$ is not a positive map for  $(-1<{\rm s}<1, |a| \neq 1)$, $({\rm s}=-1,|a|<1)$, and $({\rm s}=1,|a|>1)$.
\end{proposition}
It is interesting that all quasiprobabilities, other than possibly  
$P$ and $Q$ corresponding to ${\rm \s} \pm 1$ respectively, are on the
same footing as far as positivity of the scaling transformation 
is concerned. That Proposition \ref{prop2} {\em is silent} on $P$ (with $|a|<1$) and hence on its dual $Q$ (with $|a|>1$) is due to the fact that in the case ${\rm \s}=1$ {\em the s-ordered characteristic function of the ground state is a constant}. These two cases 
therefore need closer examination.

It turns out that help is at hand from the detailed study of the properties of  bosonic Gaussian channels and their respective operator-sum representation\cite{kraus10}, and we have the following proposition.
\begin{proposition}[Theorem 10 of \cite{kraus10}]
The scaling map $\Gamma_{-1,a}\,:  Q(\alpha;\Gamma_{-1,a}(\hr{}))= 
a^{-2} Q (a^{-1} \alpha;\hr{}),~~ a> 1$, is a trace-preserving completely positive map 
and corresponds to the quantum-limited amplifier channel ${\cal C}_2(a)$.
\label{prop3}
\end{proposition}
This leaves the final case of the scaling map $\Gamma_{1,a}$ with $a<1$ and this is addressed in the following proposition.

\begin{figure}
\centering 
\includegraphics[width=\columnwidth]{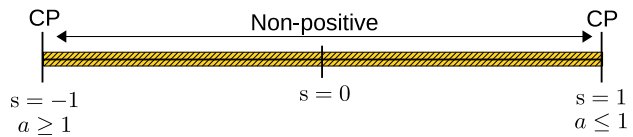}
\caption{The positivity properties of the scaling maps as depicted on the ${\rm s}$ axis. For ${\rm s} \in (-1,1)$ and $|a| \neq 1$, the scaling maps are non-positive. This is also true for $({\rm s}=-1,|a|<1$ and $({\rm s}=1,|a|>1)$. The scaling maps are completely positive for the cases $({\rm s}=-1,|a|>1)$, $({\rm s}=1,|a|<1)$, including the trivial situation of $|a|=1$ for all ${\rm s}$.  }
\label{fig5}
\end{figure}

\begin{proposition}[Theorem 6 of \cite{kraus10}]
\label{prop4}
The scaling map $\Gamma_{1,a}\,: 
P(\alpha;\Gamma_{1,a}(\hr{}))
= a^{-2} P(a^{-1} \alpha;\hr{}) $, $0< a < 1$, is a completely positive
trace-preserving map 
and corresponds to the quantum-limited attenuation channel ${\cal C}_1(a)$.
\end{proposition}
We have thus found the complete answer to the problem we set out to study\,: 
(1) for none of the ${\rm s} \in (-1,1)$  is $\sca$ a positive map
for any value of  the scale parameter $a$ other than  
the trivial values $a= \pm 1$; (2) for the special case  
${\rm \s} = -1$, the map is not positive if $|a|< 1$, and 
completely positive if $|a|\ge 1$; 
and (3) for the dual special case ${\rm \s} = 1$, the map is not positive if $|a|> 1$, and 
completely positive if $|a|\le 1$. {\em In other words, there is no 
positive but not completely positive map (entanglement witness) of the scaling 
type}. The results of this section are summarized in the following theorem.
and also depicted pictorially in Fig. \ref{fig5}. 
\begin{theorem}
Scaling maps $\sca$ are completely positive for $({\rm s} \in [-1,1],|a| = 1)$, $({\rm s}= -1, |a|>1)$, and $({\rm s} = 1, |a| <1)$. For all other values of $({\rm s},a)$ the scaling maps $\sca$ are not even positive.
\label{thm}
\end{theorem}
One may apply the scaling map followed (or preceded) by the
transpose map\,\cite{Simon} $(q,\,p)\to 
(q,\,-p)$, i.e., $\alpha \to \alpha^*$, on one subsystem of a bipartite state 
to obtain what is called the partial 
scaling map in Ref.\,\cite{Manko}. From Theorem \ref{thm} and the 
invertibility of the transpose map we have that partial scaling, with nontrivial scale 
parameter $|a| \neq 1$, is not a positive map for  any 
 ${\rm \s}\in (-1,\,1)$; for ${\rm \s} = -1$, the map is not positive if $|a|< 1$, and  
positive but not completely positive if $|a|\ge 1$; 
and for ${\rm \s} = 1$, the map is not positive if $|a|> 1$, and 
 positive but not completely positive if $|a|\le 1$. But in both the 
cases   ${\rm \s} = -1,\; |a|> 1$  
and ${\rm \s} \ = 1,\; |a|< 1$ the partial scaling map can be shown to be 
weaker than the transpose map in its capacity to witness entanglement. Further implications of Theorem 1 will be developed in the following section. 


\section{Scaling maps from the viewpoint of characteristic functions}
 
\begin{figure*}
\centering 
\includegraphics{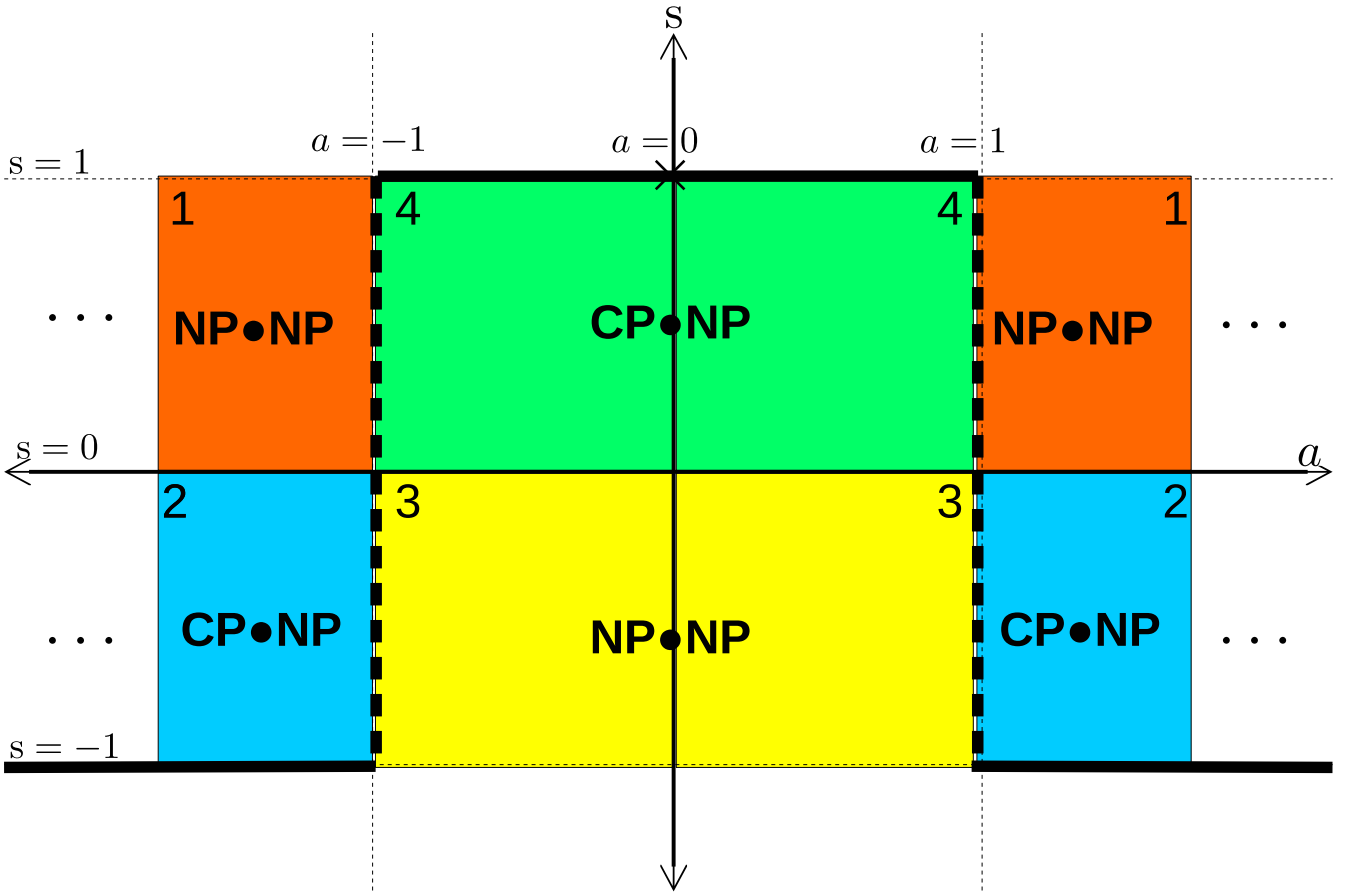}
\caption{`Phase' diagram for the scaling maps $\Gamma_{{\rm s},a}$ with respect to positivity in the $a-{\rm s}$ parameter space with ${\rm s} \in [-1,1 ]$ and $a \in {\mathbb R}$. The scaling maps corresponding to bold lines [step shape] are completely positive. The solid line at ${\rm s}=-1$ ($|a|>1$) corresponds to scaling maps that induce the scaling of the $Q$-function and the one at ${\rm s}=1$ ($|a|<1$) to that of the $P$-function, which are the quantum-limited amplifier and attenuation channels respectively. The dotted bold lines at $a=\pm 1$ correspond to unitary maps. For the rest of the parameter space the corresponding scaling maps are non-positive (regions 1 to 4). The specific reason for the failure of these maps to be positive is made transparent through its decomposition into maps that are non-positive (NP) and completely positive (CP) [the additive Gaussian classical noise channel]. However even with composition of a non-positive map with a completely positive map in regions 2 and 4, the CP map is not sufficient to render the whole scaling map positive. Finally the line $a=0$ corresponds to the pinch or constant map. Except for $(a=0,{\rm s}=1)$ which corresponds to constant output of the vacuum state (marked with an `$\times$'), the rest of the line is non-positive. The symmetry about the ${\rm s}$-axis is due to the fact that one can go from $a$ to $-a$ by a unitary operation that preserves all the properties.  } 
\label{fig1}\end{figure*}

By Theorem \ref{thm} we thus have two families of completely positive maps of the scaling type as detailed in Propositions \ref{prop3} and \ref{prop4}. These 
bosonic Gaussian channels
are traditionally described by the transformation the 
Wigner characteristic function $\chi_0$ suffers through the 
channel\,\cite{hw01,kraus10,raulrmp}. 
So, rewriting Eq. \eqref{scacar} at the level of the symmetric-ordered characteristic function (${\rm s}=0$) we have  
\begin{align}
\chi_0(\xi;\sca[\hr{}]) =  \chi_0(a\xi;\hr{}) \exp[{\rm s}(a^2-1)|\xi|^2/2 ].
\label{e14}
\end{align}
Note that we now include $a=0$ in the analysis as this can be interpreted as the constant map with a one-dimensional trivial output space. It turns out that working at the level of the symmetric-ordered characteristic functions helps to identify the reason the scaling maps fail to be positive in the corresponding parameter ranges. It is useful to introduce what we call the classical noise map given by ${\cal B}_2(b), b \in {\mathbb R}$ with its action at the level of the characteristic function given by
\begin{align}
{\cal B}_2(b) : \chi_{0}(\xi;{\cal B}_2(b) [\hr{}]) = \chi_0(\xi;\hr{}) \exp[-b|\xi|^2/2],
\label{e14b}
\end{align}
with the map being completely positive for $b \geq 0$ and non-positive for $b<0$. Note that one can replace $\chi_0$ with $\chi_s$ throughout in Eq. \eqref{e14b}  to describe the map ${\cal B}_2(b)$. In terms of the classical noise map we can rewrite the scaling maps using Eq. \eqref{e14} 
\begin{align}
\sca = {\cal B}_2({\rm s}(1-a^2)) \circ \Gamma_{0,a}.
\label{e13}
\end{align}
We now consider the case $|a| <1$.  For ${\rm s}=0$, we showed that the scaling map is not positive as can be checked by the action of the map on the vacuum state. 
For ${\rm s} < 0 $ we have that the map in Eq.\,\eqref{e14} decomposes into $\chi_0(a\xi;\hr{}) \times \exp[|{\rm s}|(1-a^2)|\xi|^2/2 ]$, i.e. a composition of two non-positive maps ${\cal B}_2(-|{\rm s}|(1-a^2)) \circ \Gamma_{0,a} $.  For the case ${\rm s} \in (0,1)$ we have that the map in Eq.\,\eqref{e14} decomposes into $\chi_0(a\xi;\hr{}) \times \exp[-{\rm s}(1-a^2)|\xi|^2/2 ]$ which is the composition of ${\cal B}_2({\rm s}(1-a^2)) \circ \Gamma_{0,a}$, a non-positive map and a completely positively map. However the product   results in a  non-positive map as can be checked on the vacuum state. In other words, the Gaussian noise term is not sufficient to compensate for the scaling of the characteristic function. The situation changes abruptly however for the case ${\rm s}=1$ when the entire map now corresponds to a completely positive map as given in Proposition\,\ref{prop4}.

Similarly, for the case $|a|>1$ we have that for ${\rm s}\geq 0$ Eq.\,\eqref{e14} reduces to 
${\cal B}_2(-{\rm s}(a^2-1)) \circ \Gamma_{0,a}$, both being non-positive maps. 
For the case ${\rm s} \in (-1,0)$ the transformation in Eq.\,\eqref{e14} can be viewed as a product of maps corresponding to 
${\cal B}_2(|{\rm s}|(a^2-1)) \circ \Gamma_{0,a}$, which is a product of a non-positive map and a completely positive map. However the combined map is non-positive due to Theorem\,\ref{thm}. Similar to the earlier case we have that at ${\rm s} =-1$ there is an abrupt transition into a completely positive map as detailed in Proposition \ref{prop3}.

So we have presented a complete description of the decomposition of the scaling maps at the level of the symmetric-ordered characteristic function. This brings out in a transparent manner the way in which the scaling maps fail to be positive in the corresponding ${\rm s}-a$ parameter space and this is depicted in Figure.\,\ref{fig1}. 
 
\subsection{Amplification and attenuation maps below complete positivity threshold}
The attenuation map is defined as ${\cal C}_1(a;b) := {\cal B}_2(b) \circ \Gamma_{0,a}, |a| <1, b>0$. It is well known that the attenuation map is completely positive for $b \geq 1-a^2$ \cite{hw01,caruso06}, entanglement breaking for $b \geq 1 + a^2$ \cite{holevoebt}, and nonclassicality breaking for $b \geq 1 + a^2$\cite{kraus10,nb,nbm,ncd}. Similarly the amplification map is defined as ${\cal C}_2(a;b) := {\cal B}_2(b) \circ \Gamma_{0,a}, |a| >1, b>0$. It is known in literature that the amplification map is completely positive for $b \geq a^2-1$ \cite{hw01,caruso06}, entanglement breaking for $b \geq 1+ a^2$ \cite{holevoebt}, and nonclassicality breaking for $b \geq 1+ a^2$ \cite{kraus10,nb,nbm,ncd}. The importance
of these two classes arise, among other things, from the fact that
{\em every noisy} amplifier or attenuator channel can be realized as a product 
of two noiseless (quantum-limited) channels that have been proved to be extremal\,\cite{kraus10,holevoext}, with one channel chosen from either of these
two classes\,\cite{kraus10,raulprl}. 

\begin{figure}
\centering 
\includegraphics[width=\columnwidth]{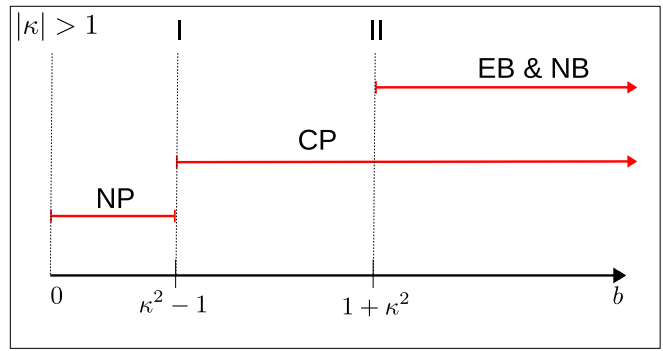}
\caption{`Phase' diagram for the amplification map ${\cal C}_2(\kappa,b)$. For a given $\kappa$ ($|\kappa|>1$), the map is non-positive for $b < \kappa^2-1$, completely positive for $b\geq \kappa^2-1$, and entanglement breaking (EB) as well as nonclassicality breaking (NB) for $b\geq 1+ \kappa^2$. In a sense phase transitions occur at $b = \kappa^2-1$ (labeled as I) and at $b = 1+\kappa^2$ (labeled as II).}
\label{fig2}
\end{figure}

\begin{figure}
\centering 
\includegraphics[width=\columnwidth]{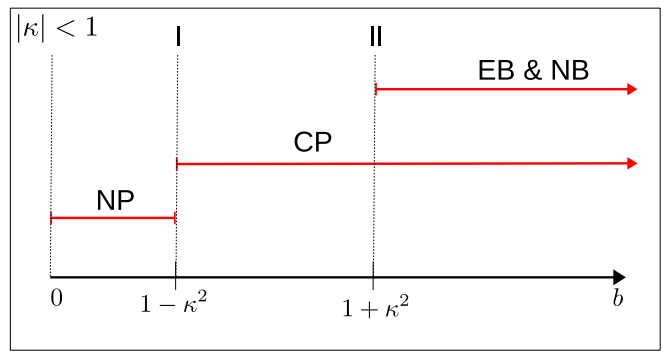}
\caption{`Phase' diagram for the attenuation map ${\cal C}_1(\kappa,b)$. For a given $\kappa$ ($|\kappa|<1$), the map is non-positive  for $b < 1-\kappa^2$, completely positive  for $b\geq 1-\kappa^2$, and entanglement breaking  as well as nonclassicality breaking  for $b\geq 1+ \kappa^2$. In a sense phase transitions occur at $b = 1-\kappa^2$ (labeled as I) and at $b = 1+\kappa^2$ (labeled as II).}
\label{fig3}
\end{figure}

Using the analysis in the preceding section, we can now complete the property of the attenuation and amplification maps for noise parameter $b$ when it is below the complete positivity threshold. We state it in the form of two theorems with respect to the amplification and attenuation maps and this is  also depicted in Figs. \ref{fig2} and \ref{fig3}. 
\begin{theorem}
The amplification map ${\cal C}_2(a;b), |a|\geq 1, b\geq 0$ is non-positive for $b < a^2-1$, completely positive for $b \geq a^2-1$, entanglement breaking and simultaneously nonclassicality-breaking for $b \geq  1+ a^2$.
\end{theorem}

\begin{theorem}
The attenuation map ${\cal C}_1(a;b), |a|\leq 1, b\geq 0$ is non-positive for $b < 1-a^2$, completely positive for $b \geq 1-a^2$, entanglement breaking and simultaneously nonclassicality-breaking for $b \geq  1+ a^2$. 
\end{theorem}

The two Gaussian families of noiseless attenuation and amplification channels are mutually
dual in multiple ways\,: (i) they are naturally described as uniform 
scaling on the dual pair of quasiprobabilities $Q$, $P$;
(ii) the physically allowed ranges for the scale parameter are mutually
reciprocal; and (iii) the Kraus operators of the two-families are mutually
dual, being simply related by hermitian conjugation.

\section{discussion}
We have presented a definitive analysis of uniform phase space scaling
maps $\sca$ with regard to positivity and summarized the result compactly in Fig. \ref{fig1} resembling a phase diagram. We find that apart from a measure-zero set of points in the $a-{\rm s}$ parameter space where the maps are complete positive, the scaling maps are non-positive. There are no scaling maps of the positive but not completely positive type, and thus cannot be used as an entanglement witness. These properties were obtained by studying the induced action of the scaling maps at the level of ${\rm s}$-ordered quasiprobabilities and a certain duality we established among the scaling maps. 

As an almost unintended fallout, we also studied the 
behavior of the amplification and attenuation maps with respect to their classical Gaussian noise parameter taken below the complete positivity threshold. For this purpose, it was useful to view the scaling maps from their induced action at the level of the characteristic functions.  
Our main finding is that below a the complete positivity threshold, the attenuation and amplification maps are non-positive. As noted earlier, this dual pair of  noiseless amplifier and attenuator 
channels are of fundamental importance due to their practical relevance as in modeling fibre optical communication.  
A natural question that follows is the behaviour of other single-mode Gaussian channels below their corresponding complete positivity or quantum-limited noise threshold. The extension to the multimode case is entirely open. 

It may be seen that nothing more interesting can be achieved by replacing the
uniform scaling by a more general scaling matrix $K$ acting on the vector
$(q, p)^{T}$. For every nonsingular $2 \times 2$ matrix $K$, there exist
symplectic matrices $S_{1}, \,S_{2} \in {\rm Sp}(2, {\mathbb R})$ such that
\begin{eqnarray}
S_{1}\,K\,S_{2} = a1\!\!1\,\,\,{\rm or}\,\,\,a\sigma_3
\end{eqnarray}
according as ${\rm det}\, K$ is positive or negative.
This simply corresponds to pre and post-processing the given
uniform scaling map by unitary (metaplectic) transformations
$U(S_1)$, $U(S_2)$; corresponding to symplectic matrices $S_{1}$, $S_{2}$. But unitary
processes do not alter positivity properties of the map. 

There is currently considerable interest in non-Gaussian states as potentially
advantageous resources in quantum information processing tasks\,\cite{robust,opat,bonifacio,mista06,anno07,grangier07,adesso,sasaki,raul12,nongpad,suyong}  and 
one measure  of non-Gaussianity based on the departure of the $Q$-distribution of a given state
from the closest Gaussian was proposed in\,\cite{non-Gaussianity}. 
The measure has the property that a Fock state $| m \rangle$ and the $m$-photon added thermal state\,\cite{PATS}
possess the same value of non-Gaussianity for every $m$, since their respective $Q$-functions
are related by a uniform scaling of the phase space variables\,\cite{non-Gaussianity} that preserves the shape of the distribution (the temperature being a monotone function of the scale parameter). There are also other possible applications to the study of nonclassicality of optical fields like in \cite{vogel10}.

We conclude with a remark on Ref.\,\cite{Manko} in respect of scaling
of the Planck constant, already referred to in the introduction. That uniform
scaling of the Wigner distribution corresponds to scaling of the Planck
constant, $\hbar \rightarrow \hbar' = a^2 \hbar$ \cite{wernerbochner}, is true\,: one can define
Wigner distributions for any chosen numerical value of the Planck constant.
Let $\Omega^{\h}$ denote the convex set of Wigner distributions corresponding 
to a chosen numerical value of the Planck constant $\hbar$. Now, expecting the scaling
map to be a positive map is to expect that the union $\Omega^{(\h_1)} \cup 
\Omega^{(\h_2)}$ of $\Omega^{(\h_1)}$ and $\Omega^{(\h_2)}$ is also a valid set
of Wigner distribution. Indeed, this expectation extends to the
convex hull of $\Omega^{(\h_1)} \cup \Omega^{(\h_2)}$, but its untenability
can be settled without going that far. Let $\h^{(1)}$ be the larger of
$\h_1$, $\h_2$. Let $W^{\h_1}(q,p;{|1 \rangle \langle 1|})$ be the Wigner distribution of
the first excited state corresponding to Planck constant $\h_1$, and let
 $W^{\h_2}(q,p;{|0 \rangle \langle 0|})$ that of the ground state corresponding
to $\h_2$. Since $W^{\h_1}(q,p;{|1 \rangle \langle 1|})$ is negative over a circle
around the origin of area proportional to $\h_1$, and since 
$W^{\h_2}(q,p;{|0 \rangle \langle 0|})$ centered at the origin is narrower than
$W^{\h_1}(q,p;{|1 \rangle \langle 1|})$, it is clear that
\begin{eqnarray}
\int dq\,dp\,W^{\h_1}(q,p;{|1 \rangle \langle 1|})\,W^{\h_2}(q,p;{|0 \rangle \langle 0|}) < 0.
\end{eqnarray}  
This shows while any numerical value of
Planck constant is acceptable, two distinct values of Planck's 
constant cannot coexist in the Wigner scheme of things! In a sense, it is fortunate that the 
Wigner scheme protects itself against a perhaps awkward question of the following type\,: if $W(q,p;\hr{})=\lambda W_1(q,p;\hr{}) + (1-\lambda) W_2(q,p;\hr{}), ~0<\lambda <1$, with $W_j(q,p;\hr{}) \in \Omega^{(\hbar_j)}$ is a Wigner distribution, to what numerical value of $\hbar$ would $W(q,p;\hr{})$ correspond to?

\begin{acknowledgments} K. K. S.  was supported by the ERC, Advanced Grant  ``IRQUAT'', Contract No. ERC-2010-AdG-267386, and Spanish MINECO FIS2013-40627-P and FIS2016-80681-P (AEI/FEDER, UE), Generalitat de Catalunya CIRIT  2014-SGR-966. K. K. S. acknowledges useful discussions with Raul Garcia-Patron. 
\end{acknowledgments}

\end{document}